# *Back to Basics?...*
or
# How can supersymmetry be used in a simple quantum cosmological model*

P.V. Moniz**

Department of Applied Mathematics and Theoretical Physics
University of Cambridge
Silver Street, Cambridge, CB3 9EW, UK

## ABSTRACT

The general theory of N=1 supergravity with supermatter is applied to a Bianchi type IX diagonal model. The supermatter is constituted by a complex scalar field and its spin-$\frac{1}{2}$ fermionic partners. The Lorentz invariant Ansatz for the wave function of the universe, $\Psi$, is taken to be as simple as possible in order to obtain *new* solutions. The wave function has a simple form when the potential energy term is set to zero. However, neither the wormhole or the Hartle-Hawking state could be found. The Ansatz for $\Psi$ used in this paper is constrasted with the more general framework of R. Graham and A. Csordás.

A quantum gravity theory constitutes a desirable goal in theoretical physics as it could lead to a proper unification of all know interactions within a quantum mechanical point of view. The inclusion of supersymmetry in quantum gravity and cosmology allowed a number of interesting results and conclusions to be achieved in the last ten years or so. Several approaches may be found in the literature, namely the triad ADM canonical formulation, the $\sigma$−model supersymmetric extension in quantum cosmology, and another approach based on Ashtekar variables (see ref. [1,2] and references therein).

The canonical quantization framework of N=1 (pure) supergravity was presented in ref. [3]. It has been pointed out that it would be sufficient, in finding a physical state, to solve the Lorentz and supersymmetry constraints of the theory due to the algebra of constraints of the theory. The presence of local supersymmetry could also contribute to the removal of divergences. Moreover, the supersymmetry constraints provide a Dirac-like square root of the second order Wheeler-DeWitt equation. Hence, we were led to solve instead a set of coupled first-order differential equations which $\Psi$ ought to satisfy.

As a result, simple forms for $\Psi$ were obtained, representing states such as the Hartle-Hawking (no-boundary) solution or the (ground state) wormhole solution [4]. Most of these results seemed then to emphasize a particular basic track, where a Dirac-like square-root structure is used to obtain directly any physical states. However, this *back to basics* line*** is not the whole story.

On the one hand, it seemed that the gravitational and gravitino modes that were allowed to be excited in each supersymmetric Bianchi model contribute in such a way as to give only very simple states or even forbid any physical solutions of the quantum constraints. Moreover, these states were found in the empty $\psi^0$ (bosonic) and fermionic filled $\psi^6$ sectors and could be either wormhole states or Hartle-Hawking no-boundary states, according to different homogeneity conditions for the gravitino field (cf. ref. [5]). However, one could *not* find both of them in the same spectrum of solutions. On the other hand, these solutions in minisuperspace were shown not to have any counterpart in the full theory because no states with finite number of fermions are possible there [6].

---

*** "Motto" proposed by the current British Prime Minister some years ago...

These results were then properly interpreted in ref. [7,8], which interested readers may wish to consult. In fact, the doubts thereby raised are entire legitimate: even though canonical quantum supergravity has more constraints than ordinary quantum gravity, it has surely more degrees of freedom than gravity. The cause for the apparent paradoxical results mentioned in the previous paragraph was the use of an Ansatz too special for the $\psi^2$ and $\psi^4$ fermionic (middle) sectors in the wave function. This Ansatz [1,2] allowed only for two bosonic amplitudes in each of the middle fermionic sectors and was constructed only from the Lorentz irreducible modes of the gravitino field. There may be 15 such invariants, when we consider the Lorentz irreducible modes of the gravitational degrees of freedom as well. However, these invariant components correspond now to a single one which satisfy a Wheeler-DeWitt type equation. For a particular factor ordering, we obtain the wormhole state in the bosonic sector and the Hartle-Hawking state in the quartic fermionic sector. Both these states are in the same spectrum of solutions but this fact has only become possible by using the Wheeler-DeWitt equation.

Hence, a Dirac-like square-root structure of gravity with local supersymmetry could not directly produce all the desired features one would like. We still had to use a Wheeler-DeWitt equation to obtain the physical quantum states altogether in a generic minisuperspace case. The (supersymmetric) Dirac-like square roots seemed not entirely sufficient, even though the Wheeler-DeWitt type equations (and Hamilton-Jacobi related equations) used in [7,8] could be obtained from the supersymmetry and Lorentz generators. Nevertheless, the overall picture that appears could be compared as considering a Dirac equation for a system with fermionic degres of freedom but that in order to obtain all physical states in the same spectrum of solutions (cf. ref. [5]) one had to turn to a Klein-Gordon type-equation. The basic track above mentioned had to be re-oriented. But let us also add the following to the discussion.

The solutions presented in [7] from the Wheeler-DeWitt type equation are not entirely new in the sense that they can be obtained using a specific definition of homogeneity conditions for the gravitinos and in the context of the "old" Ansatz construction [5]. Thus, the "old" Ansatz construction could still be usefull up to a certain point and if properly addressed.

Our objective here is to try to obtain new solutions from the "old" Ansatz construction, hoping they could bear some physical significance. We will study a locally supersymmetric (diagonal) Bianchi type-IX model coupled to a scalar supermultiplet, formed by complex spin-0 scalar fields together with their spin-$\frac{1}{2}$ partners. A set of first order differential equations obtained from the supersymmetry constraints will be used

, as the "old" procedure allows. The obtained solutions can be expected also to be present in the framework of [7,8]. The results presented here (and in ref. [9]) may give us some insight on how to generalize the approach of [7,8] to couplings with supermatter. As we will see, solutions in the top and bottom sectors as well as in some middle sectors are obtained for our model when the potential is zero (for more details see [9]). We suspect that these or similar properties and solutions would also be obtained in the "new" approach but at the expense of having to deal with more complicated Wheeler-DeWitt or Hamilton-Jacobi equations. *Of course, we could possibly find out other solutions with a more general Lorentz invariant Ansatz for the wave function of the universe.*

The Lagrangian of the general theory of N=1 supergravity with supermatter is given in eq. (25.12) of [10] and it is too long to write out here. It depends on the tetrad $e^{AA'}{}_\mu$, the spin-$\frac{3}{2}$ gravitino field $\left(\psi^A{}_\mu, \widetilde{\psi}^{A'}{}_\mu\right)$. on a vector field $A^{(a)}_\mu$ labelled by a group index $(a)$, its spin-$\frac{1}{2}$ partner $\left(\lambda^{(a)}_A, \widetilde{\lambda}^{(a)}_{A'}\right)$, a family of scalars $(\Phi^I, \Phi^{J^*})$ and their spin-$\frac{1}{2}$ partner $\left(\chi^I_A, \widetilde{\chi}^{J^*}_{A'}\right)$.

We restrict our case to a supermatter model constituted only by a scalar field and its spin$-\frac{1}{2}$ partner with a two-dimensional flat Kähler geometry. The scalar super-multiplet, consisting of a complex massive scalar field $\phi$ and spin-$\frac{1}{2}$ field $\chi, \bar{\chi}$ are chosen to be spatially homogeneous, depending only on time. The 3-metric of diagonal Bianchi type-IX is given by $h_{ij} = a_1^2 E^1{}_i E^1{}_j + a_2^2 E^2{}_i E^2{}_j + a_3^2 E^3{}_i E^3{}_j$, where $E^j{}_i$ are a basis of left-invariant one-forms on the three-sphere and $a_1, a_2, a_3$ are spatially constant. We require the components $\left(\psi^A{}_0, \overline{\psi}^{A'}{}_0\right)$ to be functions of time only. We further require that $\psi^A{}_i$ and $\tilde{\psi}^{A'}{}_i$ be spatially homogeneous in the basis $e^a{}_i$.

A quantum description can be made by studying (for example) Grassmann-algebra-valued wave functions of the form $\Psi\left[e^{AA'}{}_i, \psi^A{}_i, \bar{\chi}_A, \phi, \overline{\phi}\right]$. The choice of $\bar{\chi}_A \equiv n_A{}^{A'} \bar{\chi}_{A'}$ rather than $\chi_A$ is designed so that the quantum constraint $\overline{S}_{A'}$ should be of first order in momenta

The supersymmetry constraints become then, in differential operator form

$$\overline{S}_{A'} = -i\sqrt{2}\left[-i\hbar\frac{1}{4}\left[e_{AA'i}\psi^A_j\right]e^{CC'i}\frac{\delta}{\delta e^{CC'j}} - i\hbar\frac{1}{4}\left[e_{AA'i}\psi^A_j\right]e^{CC'j}\frac{\delta}{\delta e^{CC'i}}\right]$$

$$-\sqrt{2}\epsilon^{ijk}e_{AA'i}\,^{3(s)}\omega^A{}_{Bj}\psi^B{}_j - \frac{i}{\sqrt{2}}\hbar n_{CA'}\overline{\chi}^C\frac{\partial}{\partial\overline{\phi}}$$

$$-i\sqrt{2}\hbar h e^{K/2} P(\phi) n^A{}_{A'} e^i{}_{AB'} D^{CB'}{}_{ji}\frac{\partial}{\partial\psi^C_j} - i\sqrt{2}\hbar h e^{K/2} D_\phi P\,n^A{}_{A'}\frac{\partial}{\partial\overline{\chi}^A}$$

$$-\frac{i}{2\sqrt{2}}h^{\frac{1}{2}}\hbar\phi n^{BB'} n_{CB'}\overline{\chi}^C n_{DA'}\overline{\chi}^D\frac{\partial}{\partial\overline{\chi}^B} - \frac{i}{4\sqrt{2}}\hbar h^{\frac{1}{2}}\phi\epsilon^{ijk}e^{BB'}{}_j\psi_{kB}D^C{}_{B'li}n_{DA'}\overline{\chi}^D\frac{\partial}{\partial\psi^C_l}$$

$$-\hbar\sqrt{2}h^{\frac{1}{2}}e^B{}_{B'}{}^m n^{CB'}\psi_{mC} n_{DA'}\overline{\chi}^D\frac{\partial}{\partial\overline{\chi}^B} - \frac{i}{\sqrt{2}}\hbar\epsilon^{ijk}e_{AA'j}\psi^A_i n_{DB'}\overline{\chi}^D e^{BB'}{}_k\frac{\partial}{\partial\overline{\chi}^B}$$

$$+\frac{1}{2\sqrt{2}}\hbar h^{\frac{1}{2}}\psi_{iA}(e^B{}_{A'}{}^i n^{AC'} - e^{AC'i}n^B{}_{A'})n_{DC'}\overline{\chi}^D\frac{\partial}{\partial\overline{\chi}^B}, \tag{1}$$

while the remaining constraint is just the hermitian conjugate.

The constraints $J^{AB}\Psi = 0$, $\bar{J}^{A'B'}\Psi = 0$ imply that $\Psi$ ought to be a Lorentz-invariant function. Thus, we take expressions in which all spinor indices have been contracted together. It is reasonable also to consider only wave functions $\Psi$ which are spatial scalars, i.e., where all spatial indices $i, j, \ldots$ have also been contracted together.

We decided to construct our Lorentz invariant wave function (expressed in several fermionic sectors and corresponding bosonic amplitudes) following the view in refs. [1,2]. We are aware of its limitations (they may be particularly severe in the case of $P(\phi) \neq 0$), as far as the middle sectors are concerned. In fact, we will be neglecting Lorentz invariants built with not only the spin $\frac{1}{2}$ and $\frac{3}{2}$ Lorentz irreducible mode components of the fermionic fields but also with the ones corresponding to the gravitational degrees of freedom. We trust, however, that our results could give some indication towards a more complete implementation. Our general Lorentz-invariant wave function is then taken to be a polynomial of eight degree in Grassmann variables

$$\begin{aligned}
\Psi(a_1, a_2, a_3, \phi, \bar{\phi}) = &\ A + B_1 \beta_A \beta^A + B_2 \bar{\chi}_A \bar{\chi}^A + C_1 \gamma_{ABC} \gamma^{ABC} \\
&+ D_1 \beta_A \beta^A \gamma_{EBC} \gamma^{EBC} + D_2 \bar{\chi}_A \bar{\chi}^A \gamma_{EBC} \gamma^{EBC} + E_1 (\gamma_{ABC} \gamma^{ABC})^2 \\
&+ F_1 (\beta_A \beta^A \gamma_{EBC} \gamma^{EBC})^2 + F_2 \bar{\chi}_A \bar{\chi}^A (\gamma_{EBC} \gamma^{EBC})^2 + G_1 \beta_A \beta^A \bar{\chi}_B \bar{\chi}^B \\
&+ H_1 \beta_A \beta^A \bar{\chi}_B \bar{\chi}^B \gamma_{EDC} \gamma^{EDC} + I_1 \beta_A \beta^A \bar{\chi}_B \bar{\chi}^B (\gamma_{EDC} \gamma^{EDC})^2 \\
&+ Z_1 \bar{\chi}_A \beta^A + Z_2 \bar{\chi}_A \beta^A \gamma_{EDC} \gamma^{EDC} + Z_3 \bar{\chi}_A \beta^A (\gamma_{EDC} \gamma^{EDC})^2 \quad (2)
\end{aligned}$$

The action of the constraints operators $S_A, \bar{S}_{A'}$ on $\Psi$ lead to a system of coupled first order differential equations which the bosonic amplitude coefficients of $\Psi$ ought to satisfy. These coefficients are functions of $a_1, a_2, a_3, \phi, \bar{\phi}$.

The following table illustrate the way the quantum supersymmetry constraints operate on $\Psi$ and which types of fermionic terms can be obtained. It correspond to $\bar{S}_{A'}\Psi$. A slash over a particular bosonic amplitude means that after all the calculations have been made to simplify the corresponding equations, the expression associated with the particular coefficient is zero. A "•" means that no bosonic amplitudes in $\Psi$ can match the particular fermionic operator. As a last comment, the third and fourth columns in both tables correspond to the action of fermionic operators which appear in terms involving $P(\phi), D_\phi P(\phi)$ respectively, or their hermitian conjugates.

| ↘ | $\psi$ | $\overline{\chi}$ | $\frac{\partial}{\partial\psi}$ | $\frac{\partial}{\partial\overline{\chi}}$ | $\overline{\chi}\chi\frac{\partial}{\partial\overline{\chi}}$ | $\psi\overline{\chi}\frac{\partial}{\partial\psi}$ | $\psi\overline{\chi}\frac{\partial}{\partial\overline{\chi}}$ |
|---|---|---|---|---|---|---|---|
| $\beta$ | $A$ | • | $B_1$ | $Z_1$ | • | • | • |
| $\gamma$ | $A$ | • | $\not{C}_1$ | • | • | • | • |
| $\overline{\chi}$ | • | $A$ | $Z_1$ | $B_2$ | • | • | • |
| $\beta\beta\overline{\chi}$ | $Z_1$ | $B_1$ | • | $G_1$ | • | $B_1$ | $Z_1$ |
| $\beta\beta\gamma$ | $B_1$ | • | $\not{D}_1$ | • | • | • | • |
| $\beta\overline{\chi}\chi$ | $B_2$ | $Z_1$ | $G_1$ | • | $Z_1$ | $Z_1$ | $B_2$ |
| $\beta\overline{\chi}\gamma$ | $Z_1$ | • | • | • | • | $C_1, B_1$ | $Z_1$ |
| $\beta\gamma\gamma$ | $C_1$ | • | $D_1$ | $Z_2$ | • | • | • |
| $\overline{\chi}\chi\gamma$ | $B_2$ | • | $\not{D}_2$ | • | • | $\not{Z}_1$ | $\not{B}_2$ |
| $\overline{\chi}\gamma\gamma$ | • | $G_1$ | $Z_2$ | $B_2$ | • | $C_1$ | • |
| $\gamma\gamma\gamma$ | $C_1$ | • | $\not{E}_1$ | • | • | • | • |
| $\beta(\gamma\gamma)^2$ | $E_1$ | • | $F_1$ | $Z_3$ | • | • | • |
| $\gamma\beta\beta\gamma\gamma$ | $D_1$ | • | $\not{F}_1$ | • | • | • | • |
| $\gamma\overline{\chi}\chi\gamma\gamma$ | $D_2$ | • | $\not{F}_2$ | • | • | $Z_2$ | $\not{D}_2$ |
| $\beta\overline{\chi}\chi\gamma\gamma$ | $D_2$ | $Z_2$ | $H_1$ | • | $Z_2$ | $Z_2$ | $D_2$ |
| $\gamma\beta\beta\overline{\chi}\chi$ | $G_1$ | • | $\not{H}_1$ | • | • | $Z_2$ | $\not{H}_1$ |
| $\overline{\chi}(\gamma\gamma)^2$ | • | $E_1$ | $Z_3$ | $F_2$ | • | $E_1$ | • |
| $\overline{\chi}\beta\gamma\gamma\gamma$ | $Z_2$ | • | • | • | • | $D_1, E_1$ | $Z_2$ |
| $\overline{\chi}\beta\beta\gamma\gamma$ | $Z_2$ | $D_1$ | • | $H_1$ | • | $D_1$ | $Z_2$ |
| $\beta\overline{\chi}\chi(\gamma\gamma)^2$ | $F_2$ | $Z_3$ | $I_1$ | • | $Z_3$ | $Z_3$ | $F_2$ |
| $\gamma\beta\beta\overline{\chi}\chi\gamma\gamma$ | $H_1$ | • | $\not{I}_1$ | • | • | • | $\not{H}_1$ |
| $\overline{\chi}\beta\beta(\gamma\gamma)^2$ | $Z_3$ | $F_1$ | • | $I_1$ | • | $F_1$ | $Z_3$ |

**Table 1**: Action of $\overline{S}_{A'}$ on $\Psi$

We then analysed two possible cases, namely when the scalar field dependent potential in the supermatter content was either arbitrary or identically set to zero. In the former, our main (and perhaps, solely) conclusion was that the adequacy of our Ansatz for $\Psi$ is severly limited. The constraints imply $\Psi = 0$ for some simplifying assumptions. In a more general setting, no easy way is apparent of obtaining an analytical solution for the full set of equations. In fact, exponential terms as $e^{\phi\overline{\phi}}$ lead to serious difficulties.

In the latter, however, we found out that $\Psi$ had indeed a very simple form. More precisely, we obtained

$$\Psi = f(\overline{\phi})e^{[-a_1^2-a_2^2-a_3^2]} + h(\phi)a_1a_2a_3e^{[-a_1^2-a_2^2-a_3^2]}e^{-2\pi^2\phi\overline{\phi}}\overline{\chi}_A\overline{\chi}^A$$

$$+ g(\overline{\phi})a_1a_2a_3e^{[a_1^2+a_2^2+a_3^2]}e^{-2\pi^2\phi\overline{\phi}}\beta_A\beta^A(\gamma_{BCD}\gamma^{BCD})^2 + k(\phi)e^{[a_1^2+a_2^2+a_3^2]}\overline{\chi}_A\overline{\chi}^A\beta_E\beta^E(\gamma_{BCD}\gamma^{BCD})^2. \quad (3)$$

Unfortunately, we cannot find the Hartle-Hawking and wormhole states in eq.(3) [9].

It seems though from our analysis and discussion hereby presented that when supermatter is present, one cannot simply expect to follow a somewhat simple relation between the old and new frameworks. Perhaps the approach in [7,8] properly applied to our Bianchi-IX model could be able to find out the Hartle-Hawking states in *other* middle sectors. However, the absence of a wormhole state is another issue to be properly addressed [10]. Overall, we hope our results (see ref. [9] as well) could provide new ground for a debate on supersymmetric quantum cosmology with supermatter fields.

## ACKNOWLEDGEMENTS

The author is thankful to R. Caldwell for assistance with TeX macros, to A.D.Y. Cheng for many discussions and for sharing his points of view and to S.W. Hawking for helpful conversations. Discussions with O. Obregon and R. Graham are also acnowledged. The author gratefully acknowledges the Instituto de Fisica de la Universidad de Guanajuato, Mexico, for supporting part of his 3 weeks visit as well as the support of a Human Capital and Mobility (HCM) Fellowship from the European Union (Contract ERBCHBICT930781).